\documentclass[prd,twocolumn]{revtex4}

\usepackage{graphicx}
\usepackage{dcolumn}
\usepackage{amsmath}
\usepackage{epsfig}
\usepackage{amsfonts}
\usepackage{amssymb}
\usepackage{revsymb}

\newcommand{\black}[1]{\mathbf{#1}}

\makeatother

\begin{document}


\title{Cosmic Topology: a Brief Overview}

\author{M.J. Rebou\c{c}as}
\email{reboucas@cbpf.br}
\affiliation{Centro Brasileiro de Pesquisas F\'{\i}sicas\\
Departamento de Relatividade e Part\'{\i}culas \\
Rua Dr.\ Xavier Sigaud, 150 , 22290-180 Rio de Janeiro -- RJ, Brazil}

\author{G.I. Gomero}
\email{german@ift.unesp.br}
\affiliation{Instituto de F\'{\i}sica Te\'orica \\
Universidade Estadual Paulista \\
Rua Pamplona, 145, 01405-900 S\~ao Paulo -- SP, Brazil \\}

\date{\today}

\begin{abstract}
Questions such as whether we live in a spatially finite universe, and 
what its shape and size may be, are among the fundamental open problems 
that high precision modern cosmology needs to resolve. These questions
go beyond the scope of general relativity (GR), since as a (local) 
metrical theory GR leaves the global topology of the universe 
undetermined.  Despite our present-day inability to \emph{predict\/} 
the topology of the universe, given the wealth of increasingly accurate
astro-cosmological observations it is expected that we should be able
to \emph{detect} it.
An overview of basic features of cosmic topology, the main methods for its 
detection, and observational constraints on detectability are briefly 
presented. Recent theoretical and observational results related to cosmic 
topology are also discussed.
\end{abstract}


\maketitle

\section{Introduction} 
\label{sec:intro}

Is the space where we live finite or infinite? The popular ancient 
Greek finite-world response, widely accepted in medieval Europe, is 
at a first sight open to a devastating objection: in being finite the 
world must have a limiting boundary. But this is impossible, because 
a boundary can only separate one part of the space from another: why 
not redefine the universe to include that other part? In this way a 
common-sense response to the above old cosmological question is that 
the universe has to be infinite otherwise something else would have to 
exist beyond its limits.  This answer seems to be obvious and needing 
no further proof or explanation. However, in mathematics it is known 
that there are compact spaces (finite) with no boundary. They are 
called closed spaces. Therefore, our universe can well be spatially 
closed (topologically) with nothing else beyond its 'spatial limits'. 
This may be difficult to visualize because we are used to viewing from 
'outside' objects which are embedded in our regular $3$--dimensional 
space. But there is no need to exist any region beyond the spatial 
extent of the universe.  

Of course, one might still ask what is outside such a closed 
universe. But the underlying assumption behind this question 
is that the ultimate physical reality is an infinite Euclidean 
space of some dimension, and nature needs not to adhere to 
this theoretical embedding framework. It is perfectly acceptable 
for our $3$--space not to be embedded in any higher-dimensional 
space with no physical grounds. 

Whether the universe is spatially finite and what is its size and 
shape are among the fundamental open problems that high precision 
modern cosmology seeks to resolve. These questions of topological 
nature have become particularly topical, given the wealth of 
increasingly accurate astro-cosmological observations, especially
the recent observations of the cosmic microwave background radiation
(CMBR)~\cite{WMAP}. An important point in the search for answers
to these questions is that as a (local) metrical theory general 
relativity (GR) leaves the global topology of the universe 
undetermined. Despite this inability to \emph{predict\/} the topology 
of the universe we should be able to devise strategies and methods
to \emph{detect} it by using data from astro-cosmological observations.

The aim of the article is to give a brief review of the main points 
on cosmic topology addressed in the talk delivered by one of us (MJR) 
in the XXIV Brazilian National Meeting on Particles and Fields, and 
discuss some recent results in the field. 
The outline of our paper is as follows. In section~\ref{sec:origin}
we discuss how the cosmic topology issue arises in the context of 
the standard cosmology, and what are the main observational consequences 
of a nontrivial topology for the spatial section of the universe. 
In section~\ref{sec:statmethods} we review the two main statistical 
methods to detect cosmic topology from the distribution of discrete 
cosmic sources.
In section~\ref{sec:CinSky} we describe the search for circles 
in the sky, an important method which has been devised for the 
detection of cosmic topology from CMBR.
In section~\ref{sec:detect} we discuss the detectability of cosmic 
topology and present examples on how one can decide whether a 
given topology is detectable or not according to recent 
observations.
Finally, in section~\ref{sec:news+remarks} we briefly discuss  
recent results on cosmic topology, and present some concluding 
remarks.

\section{Nontrivial topology and physical consequences} 
\label{sec:origin}

The isotropic expansion of the universe, the primordial abundance of 
light elements and the nearly uniform cosmic microwave background radiation 
constitute the main observational pillars for the standard cosmological 
model, which provides a very successful description of the universe. 
Within the framework of standard cosmology, the universe is described 
by a space-time manifold $\mathcal{M}_4 = \mathbb{R} \times M$ endowed 
with the homogeneous and isotropic Robertson-Walker (RW) 
metric
\begin{equation}
\label{RWmetric}
ds^2 = - c^2dt^2 + R^2(t)\,\{\, d\chi^2 + f^2(\chi)\,
       [\,d\theta^2 + \sin^2\theta\,d\phi^2\,] \,\} \;,
\end{equation}
where $t$ is a cosmic time, $f(\chi) = (\chi, \sin\chi,\sinh\chi)$ 
depending on the sign of the constant spatial curvature $k=(0, 1, -1)$, 
and $R(t)$ is the scale factor.
The spatial section $M$ is often taken to be one of the following 
(simply-connected) spaces: Euclidean $\mathbb{E}^3$, spherical 
$\mathbb{S}^3$, or hyperbolic space $\mathbb{H}^3$.
This has led to a common misconception that the Gaussian curvature 
$k$ of $M$ is all one needs to decide whether this $3$--space
is finite or not.  
However, the $3$-space $M$ may equally well be one of the possible 
quotient manifolds $M = \widetilde{M}/\Gamma$, where $\Gamma$ is 
a discrete and fixed-point free group of isometries of the 
corresponding covering space $\widetilde{M} = (\mathbb{E}^3, 
\mathbb{S}^3, \mathbb{H}^3)$. 
Quotient manifolds are multiply connected: compact in three 
independent directions with no boundary (closed), or compact
in two or at least one independent direction. 
The action of $\Gamma$ tessellates $\widetilde{M}$ into identical 
cells or domains which are copies of what is known as fundamental 
polyhedron (FP). In forming the quotient manifolds $M$ the 
essential point is that they are obtained from $\widetilde{M}$ 
by identifying points which are equivalent under the action of 
the discrete group $\Gamma$. Hence, each point on the quotient 
manifold $M$ represents all the equivalent points on the 
covering manifold $\widetilde{M}$.
A simple example of quotient manifold in two dimensions is the
$2$--torus $T^2 = S^1 \times S^1= \mathbb{E}^2/\Gamma$. The 
covering space clearly is $\mathbb{E}^2$, and a FP is a 
rectangle with opposite sides identified. This FP tiles the
covering space $\mathbb{E}^2$. The group $\Gamma$ consists
of discrete translations associated with the side identifications.

In a multiply connected space any two points can always be joined 
by more than one geodesic. Since the radiation emitted by cosmic 
sources follows geodesics, the immediate observational consequence 
of a spatially closed universe is that light from distant objects 
can reach a given observer along more than one path --- the sky 
may show multiple images of radiating sources [cosmic objects or 
cosmic microwave background radiation from the last scattering 
surface - (LSS)]. Clearly we are assuming here that the radiation 
(light) must have sufficient time to reach the observer at $p \in M$
(say) from multiple directions, or put in another way, that the universe 
is sufficiently small so that this repetitions can be observed. In 
this case the observable horizon $\chi_{hor}$  exceeds at least 
the smallest characteristic size of $M$  at $p\,$ 
\footnote{This is the so-called \emph{injectivity radius\/} 
$r_{inj}(p)$. A more detailed discussion on this point will 
be given in section~\ref{sec:detect}.},
and the topology of the universe is in principle detectable.

A question that arises at this point is whether one can use the 
topological multiple images of the same celestial objects such 
as cluster of galaxies, for example, to determine a nontrivial cosmic 
topology 
\footnote{There are basically three types of catalogues which can 
possibly be used in the search for multiple images in the universe: 
clusters of galaxies, with redshifts up to $z_{max} \approx 0.3$; 
active galactic nuclei with a redshift cut-off of $z_{max} \approx 4$; 
and maps of the CMBR with a redshift of $z \approx 10^3$.}.
Besides the pioneering work by Ellis~\cite{Ellis1971},
others including Sokolov and Shvartsman~\cite{SokolovShvartsman1974}, 
Fang and Sato~\cite{FangSato1985}, Starobinskii~\cite{SokolovStarobinsky1975}, 
Gott~\cite{Gott1980} and Fagundes~\cite{Fagundes1983} and Fagundes
and Wichoski~\cite{FagundesWichoski1987}, used this feature in 
connection with closed flat and non-flat universes. It has been
recently shown that the topology of a closed flat universe can 
be reconstructed with the observation of a very small 
number of multiple images~\cite{Gomero2003a}. 

In practice, however, the identification of multiple images is
a formidable observational task to carry out because it involves 
a number of problems, some of which are:
\begin{itemize}
\item
Images are seen from different angles (directions), which makes it 
very hard to recognize them as identical due to morphological 
effects;
\item
High obscuration regions or some other object can mask or even hide
the images;
\item
Two images of a given cosmic object at different distances correspond 
to different periods of its life, and so they are in different stages 
of their evolutions, rendering problematic their identification as
multiple images.
\end{itemize}

These difficulties make clear that a direct search for multiples images is
not overly promising, at least with available present-day technology. 
On the other hand, they motivate new search strategies and methods to 
determine (or just detect) the cosmic topology from observations. 
In the next section we shall discuss statistical methods, which have 
been devised to determine a possible nontrivial topology of the 
universe from the distribution of discrete cosmic sources.

\section{Pair Separations Statistical methods} 
\label{sec:statmethods}

On the one hand the most fundamental consequence of a multiply connected
spatial section $M$ for the universe is the existence of multiple 
images of cosmic sources, on the other hand a number of observational 
problems render the direct identification of these images practically 
impossible. 
In the statistical approaches we shall discuss in this section instead of 
focusing on the direct recognition of multiple images, one treats 
statistically the images of a given cosmic source, and use (statistical) 
indicators or signatures in the search for a sign of a nontrivial 
topology. Hence the statistical methods are not plagued by direct 
recognition difficulties such as morphological effects, and distinct 
stages of the evolution of cosmic sources.
 
The key point of these methods is that in a universe with detectable 
nontrivial topology at least one of the characteristic sizes
of the space section $M$ is smaller than a given  
survey depth $\chi_{obs}$, so the sky should show multiple images of
sources, whose $3$--D positions are correlated by the isometries  
of the covering group $\Gamma$. These methods rely on the fact that 
the correlations among the positions of these images can be couched 
in terms of distance correlations between the images, and use 
statistical indicators to find out signs of a possible nontrivial 
topology of $M$. 

In 1996 Lehoucq \emph{et al.\/}~\cite{LeLaLu1996} proposed the 
first statistical method (often referred to as cosmic crystallography), 
which looks for these correlations by using pair separations histograms 
(PSH). To build a PSH we simply evaluate a suitable one-to-one function 
$F$ of the distance $d$ between a pair of images in a catalogue $\mathcal{C}$, 
and define $F(d)$ as the pair separation: $s = F(d)$. Then we depict 
the number of pairs whose separation lie within certain sub-intervals $J_i$ 
partitions of $( 0,\,s_{max} ]$, where $s_{max} = F(2\chi_{max})$, and 
$\chi_{max}$ is the survey depth of $\mathcal{C}$. A PSH is just a 
normalized plot of this counting. In most applications in the 
literature the separation is taken to be simply the distance between 
the pair $s=d$ or its square $s=d^2$, $J_i$ being, respectively, a 
partition of $(0, 2 \chi_{max}]$ and $(0, 4^{}_{} \chi_{max}^2]$. 

The PSH building procedure can be formalized as follows. Consider a
catalogue $\mathcal{C}$ with $n$ cosmic sources and denote by
$\eta(s)$ the number of pairs of sources whose separation is $s$.
Divide the interval $(0,s_{max}]$ in $m$ equal sub-intervals
(bins) of length    $ \delta s = s_{max} / {m}$, being        
\[
J_i = (s_i - \frac{\delta s}{2} \, , \, s_i + \frac{\delta
s}{2}] \;,; \quad  i=1,2, \dots ,m \;\,, \quad
\]
and centered at $ s_i = \,(i - \frac{1}{2})\, \delta s \,$.
The PSH is defined as the following counting function: 
\begin{equation}
\label{PSH}
\Phi(s_i)=\frac{2}{n(n-1)}\,\,\frac{1}{\delta s}\,
               \sum_{s \in J_i} \eta(s) \; ,
\end{equation}
which can be seen to be subject to the normalization condition
$\sum_{i=1}^m \Phi(s_i)\,\, \delta s = 1 \;.$
An important advantage of using \emph{normalized\/} PSH's is that
one can compare histograms built up from catalogues with
dif\/ferent number of sources. 

\begin{figure}[thb] 
\includegraphics{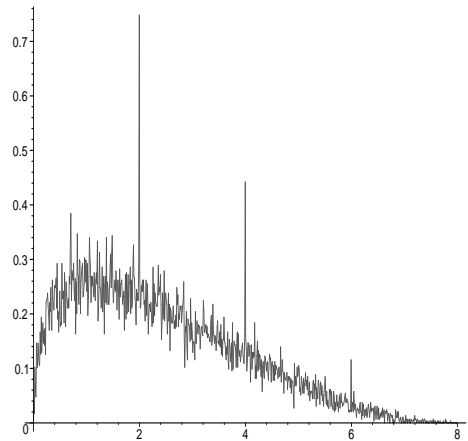}
\caption{\label{fig:PSH-T3} Typical PSH for a flat universe with
a $3$--torus topology. The horizontal axis gives the squared pair
separation $s^2$, while the vertical axis provides a normalized
number of pairs.} 
\end{figure}

An example of PSH obtained through simulation for a universe 
with nontrivial topology is given in Fig.~\ref{fig:PSH-T3}. 
Two important features should be noticed: (i) the presence of 
the very sharp peaks (called spikes); and (ii) the existence of a
'mean curve' above which the spikes stands. This curve 
corresponds to an expected pair separation histogram (EPSH) 
$\Phi_{exp}(s_i)$, which is a typical PSH from which the 
statistical noise has been withdrawn, that is 
$\Phi_{exp}(s_i) = \Phi(s_i)  - \rho(s_i)\,$, where $\rho(s_i)$ 
represents the statistical f\/luctuation that arises in the 
PSH $\Phi(s_i)$.  
 
The primary expectation was that the distance correlations 
would manifest as topological spikes in PSH's, and that 
the spike spectrum of topological origin would be a definite 
signature of the topology~\cite{LeLaLu1996}. 
While the first simulations carried out for specific flat manifolds 
appeared to confirm this expectation~\cite{LeLaLu1996}, 
histograms subsequently generated for specific hyperbolic 
manifolds revealed that the corresponding PSH's exhibit no spikes%
~\cite{LeLuUz1999,FagGaus1998}.
Concomitantly, a theoretical statistical analysis of the distance 
correlations in PSH's was accomplished, and a proof was presented 
that the spikes of topological origin in PSH's are due to just one 
type of isometry: the Clifford translations (CT)~\cite{GTRB1998}, which 
are isometries $g_t \in \Gamma $ such that for all $p \in \widetilde{M}$ 
the distance $d(p, g_t p)$ is a constant (see also in this regard%
~\cite{LeLuUz1999}). Clearly the CT's reduce to the regular translations 
in the Euclidean spaces (for more details and simulations see~\cite{GRT2000a,GRT2000b,GRT2001a}).
Since there is no CT translation in hyperbolic geometry this 
result explains the absence of spikes in the PSH's of hyperbolic 
universes with nontrivial detectable topology. On the other hand, 
it also makes clear that distinct manifolds which admit the same 
Clifford translations in their covering groups present the same spike
spectrum of topological origin. Therefore the topological spikes 
are not sufficient for unambiguously determine the topology of the 
universe.

In spite of these limitations, the most striking evidence of 
multiply-connectedness in PSH's is indeed the presence of topological 
spikes, which result from translational isometries $g_t \in \Gamma\,$.
It was demonstrated~\cite{GRT2000a,GTRB1998} that the other isometries $g$ 
manifest as very tiny deformations of the expected pair separation 
histogram $\Phi^{sc}_{exp}(s_i)$ corresponding to the underlying 
simply connected universe~\cite{BernuiTeixeira1999,Reboucas2000}.
Furthermore, in PSH's of universes with nontrivial topology the 
amplitude of the sign of non-translational isometries was shown to be
smaller than the statistical noise~\cite{GRT2000a}, making clear
that one cannot use PSH to reveal these isometries.

In brief, the only significant (measurable) sign of a nontrivial 
topology in PSH are the spikes, but they can be used merely to 
disclose (not to determine) a possible nontrivial topology 
of universes that admit Clifford translations: any flat, some 
spherical, and no hyperbolic universes. 

The impossibility of using the PSH method for the detection of 
the topology of hyperbolic universes motivated the development 
of a new scheme called \emph{collecting correlated pairs method} 
(CCP method)~\cite{UzLeLu1999} to search for cosmic topology. 

In the CCP method it is used the basic feature of the isometries, 
i.e., that they preserve the distances between pairs of images.
Thus, if $(p,q)$ is a pair of arbitrary images (correlated or not)
in a given catalogue $\mathcal{C}$, then for each $g \in \Gamma$ 
such that the pair $(gp,gq)$ is also in $\mathcal{C}$ we obviously 
have
\begin{equation}   \label{typeI}
d(p,q) = d(gp,gq) \; .
\end{equation}
This means that for a given (arbitrary) pair $(p,q)$ of images 
in $\mathcal{C}$, if there are $n$ isometries $g \in \Gamma$ such 
that both images $gp$ and $gq$ are still in $\mathcal{C}$, then
the separation $s(p,q)$ will occur $n$ times. 

The easiest way to understand the CCP method is by looking into  
its computer-aimed procedure steps, and then examine the consequences 
of having a multiply connected universe with detectable topology. 
To this end, let $\mathcal{C}$ be a catalogue with $n$ sources, 
so that one has $P = n(n-1)/2$ pairs of sources. The CCP procedure 
consists on the following steps:
\begin{enumerate}
\item 
Compute the $P$ separations $s(p,q)$, where $p$ and $q$ are two images 
in the catalogue $\mathcal{C}$;
\item 
Order the $P$ separations in a list $\{s_i\}_{1 \leq i \leq P}$ such 
that $s_i \leq s_{i+1} \,$; 
\item 
Create a list of \emph{increments} $\{ \Delta_i \}_{1 \leq i \leq P-1}$, 
where $\Delta_i = s_{i+1} - s_i \,$;.
\item 
Def\/ine the CCP index as
\begin{displaymath}
\mathcal{R} = \frac{\mathcal{N}}{P-1} \, ,
\end{displaymath}
where $\mathcal{N} = Card\{i \,:\, \Delta_i = 0\}$ is the
number of times the increment is null.
\end{enumerate}

If the smallest characteristic length of $M$ exceeds the survey 
depth ($r_{inj} > \chi_{obs}$) the probability that two pairs of 
images are separated by the same distance is zero, so 
$\mathcal{R} \approx 0$. On the other hand, in a universe with
detectable nontrivial topology ($\chi_{obs}> r_{inj}$) given 
$g \in \Gamma$, if $p$ and $q$ as well as $gp$ and $gq$ are 
images in $\mathcal{C}$, then: 
(i) the pairs $(p,q)$ and $(gp,gq)$ are separated by the same 
distance; and 
(ii) when $\Gamma$ admits a translation $g_t$ the pairs $(p,g_t p)$ 
and $(q,g_t q)$ are also separated by the same distance.
It follows that when a nontrivial topology is detectable, and
a given catalogue $\mathcal{C}$ contains multiple images, then 
$\mathcal{R} > 0$, so the CCP index is an indicator of a 
detectable nontrivial topology of the spatial section $M$ of 
the universe. Note that although $\mathcal{R}> 0$ can be used 
as a sign of multiply connectedness, it gives no indication as 
to what the actual topology of $M$ is. Clearly whether one 
can find out that $M$ is multiply connected (compact in at 
least one direction) is undoubtedly a very important step, 
though. 

In more realistic situations, uncertainties in the determination
of positions and separations of images of cosmic sources are 
dealt with through the following extension of the CCP index: 
\begin{displaymath}
\mathcal{R_{\epsilon}} = \frac{\mathcal{N_{\epsilon}}}{P-1} \; ,
\end{displaymath}
where $\mathcal{N_{\epsilon}} = Card\{i \, : \, \Delta_i \leq
\epsilon\}$, and $\epsilon > 0$ is a parameter that quantifies
the uncertainties in the determination of the pairs separations.

Both PSH and CCP statistical methods rely on the accurate 
knowledge of the three-dimensional positions of the cosmic 
sources. The determination of these positions, however, 
involves inevitable uncertainties, which basically arises 
from:
(i) uncertainties in the determination of the values of the
cosmological density parameters $\Omega_{m0}$ and
$\Omega_{\Lambda 0}$; 
(ii) uncertainties in the determination
of both the red-shifts (due to spectroscopic limitations),
and the angular positions of cosmic objects (displacement,
due to gravitational lensing by large scale objects, e.g.); and 
(iii) uncertainties due to the peculiar velocities of 
cosmic sources, which introduce peculiar red-shift 
corrections.
Furthermore, in most studies related to these methods the catalogues 
are taken to be complete, but real catalogues are incomplete: objects 
are missing due to selection rules, and also most surveys are not 
full sky coverage surveys.
Another very important point to be considered regarding these
statistical methods is that most of cosmic objects do not have 
very long lifetimes, so there may not even exist images of a
given source at large red-shift. This poses the important problem 
of what is the suitable source (candle) to be used in these 
methods. 

Some of the above uncertainties, problems and limits of the 
statistical methods have been discussed by Lehoucq \emph{et al.\/}%
~\cite{LeUzLu2000}, but the robustness of these methods still 
deserves further investigation. So, for example, a quantitative 
study of the sensitivity of spikes and CCP index with respect to 
the uncertainties in the positions of the cosmic sources, which 
arise from unavoidable uncertainties in values of the density 
parameters is being carried out~\cite{BeGoMoRe2004}. 
In~\cite{BeGoMoRe2004} it is  also determined the  optimal 
values of the bin size (in the PSH method) and the $\epsilon$ 
parameter (in the CCP method) so that the correlated pairs are 
collected in a way that the topological sign is preserved.

For completeness we mention that Bernui~\cite{Bernui2003}
has worked with a similar method which uses angular pair separation
histogram (APSH) in connection with CMBR.

To close this section we refer the reader to references%
~\cite{RoukemaEdge1997,FagGaus1999}, which present variant 
statistical methods (see also the review article~\cite{Reviews}).
 
\section{Circles in the sky} 
\label{sec:CinSky}

The deepest surveys currently available are the CMBR temperature anisotropy 
maps with $z_{LSS} \approx 10^3$. Thus, given the current 
high quality and resolution of such maps, the most promising searches 
for cosmic topology through multiple images of radiating sources are 
based on pattern repetitions of these CMBR anisotropies. 
 
The last scattering surface (LSS) is a sphere of radius $\chi_{LSS}$ on 
the universal covering manifold of the comoving space at present time. 
If a nontrivial topology of space is detectable, then this sphere 
intersects some of its topological images. Since the intersection of 
two spheres is a circle, then CMBR temperature anisotropy maps will 
have matched circles, i.e. pairs of equal radii circles (centered on 
different point on the LSS sphere) that have the same pattern of 
temperature variations~\cite{CSS1998}.

These matched circles will exist in CMBR anisotropy maps of universes 
with any detectable nontrivial topology, regardless of its geometry. Thus 
in principle the search for `circles in the sky' can be performed without 
any  \emph{a priori} information (or assumption) on the geometry, and on 
the topology of the universe. 

The mapping from the last scattering surface to the night sky sphere is
a conformal map. Since conformal maps preserves angles, the identified circle
at the LSS would appear as identified circles on the night sky sphere.
A pair of matched circles is described as a point in a six-dimensional 
parameter space. These parameters are the centers of each circle, which are 
two points on the unit sphere (four parameters), the angular radius of both 
circles (one parameter), and the relative phase between them (one parameter). 

Pairs of matched circles may be hidden in the CMBR maps if the universe has
a detectable topology. Therefore to observationally probe nontrivial topology
on the available largest scale, one needs a statistical approach to scan all-sky 
CMBR maps in order to draw the correlated circles out of them. To this 
end, let $\black{n}_1 = (\theta_1 , \varphi_1)$ and $\black{n}_2 = (\theta_2 , 
\varphi_2)$ be the center of two circles $C_1$ and $C_2$ with angular radius 
$\nu$. The search for the matching circles can be performed by computing the 
following correlation function~\cite{CSS1998}:
\begin{equation}
\label{CirSky}
S(\alpha) = \frac{\langle 2 T_1(\pm \phi) T_2(\phi + \alpha) 
\rangle}{\langle T_1(\pm \phi)^2 + T_2(\phi + \alpha)^2 \rangle} 
\; ,
\end{equation}
where $T_1$ and $T_2$ are the temperature anisotropies along 
each circle, $\alpha$ is the relative phase between the two circles, 
and the mean is taken over the circle parameter $\phi\,$: 
$\langle \;\, \rangle = \int_{0}^{2\pi} d\phi$. The plus $(+)$ and
minus $(-)$ signs in (\ref{CirSky}) correspond to circles correlated, 
respectively, by non-orientable and orientable isometries.
 
For a pair of circles correlated by an isometry (perfectly matched) 
one has $T_1(\pm \phi) = T_2(\phi + \alpha_*)$ for some 
$\alpha_*$, which gives $S(\alpha_*) = 1$, otherwise the circles 
are uncorrelated and so $S(\alpha) \approx 0$. 
Thus a peaked correlation function around some $\alpha_*$ would 
mean that two matched circles, with centers at $\black{n}_1$ and $
\black{n}_2$, and angular radius $\nu$, have been detected. 

{}From the above discussion it is clear that a full search for matched 
circles requires the computation of $S(\alpha)$, for any permitted $\alpha$, 
sweeping the parameter sub-space $(\theta_1 , \varphi_1, \theta_2 , \varphi_2, 
\nu)$, and so it is indeed computationally expensive. Nevertheless, such 
a search is currently in progress, and preliminary results 
using the first year WMAP data indicate the lack of antipodal, and 
nearly antipodal, matched circles with radii larger than $25^\circ$%
~\cite{CSSK2003}. Here nearly antipodal means circles whose center are 
separated by more than $170^\circ$.

According to these first results, the possibility that our universe has a 
torus-type \emph{local shape\/} is discarded, i.e. any flat topology with 
translations smaller than the diameter of the sphere of last scattering 
is ruled out. As a matter of fact, as they stand these preliminary results 
exclude any topology whose isometries produce antipodal images of the observer, 
as for example the Poincar\'e dodecahedron model~\cite{Poincare}, or any 
other homogeneous spherical space with detectable isometries. 

Furthermore, since detectable topologies (isometries) do not produce,
in general, antipodal correlated circles, a little more can be inferred from
the lack or nearly antipodal matched circles. Thus, in a flat universe,
e.g., any screw motion may generate pairs of circles that are not even 
nearly antipodal, provided that the observer's position is far enough 
from the axis of rotation~\cite{Gomero2003b}. As a consequence, our 
universe can still have a flat topology, other than the $3$-torus, 
but in this case the axis of rotation of the screw motion corresponding 
to a pair of matched circles would pass far from our position. 
Similar results also hold for spherical universes with non-translational 
isometries generating pairs of matched circles. Indeed, the universe could 
have the topology of, e.g., an inhomogeneous lens space $L(p,q)$, but with 
both equators of minimal injectivity radius passing far from us 
\footnote{
In spherical geometry, the equators of minimal injectivity radius of an 
orientable non-translational isometry correspond to the axis of rotation 
of an Euclidean screw motion~\cite{MGRT2003a}.}.
These points also make clear the crucial importance of the position of 
the observer relative to the 'axis of rotation' in the matching circles
search scheme for inhomogeneous spaces (in this regard see also%
~\cite{RiazUzLeLu2003}).

To conclude, `circles in the sky' is a promising method in the search for 
the topology of the universe, and may provide more general and 
realistic constraints on the shape and size of our universe in the near 
future. An important point in this regard is the lack of computational
less expensive search for matched circles, which can be archived by
restricting (in the light of observations) the expected detectable 
isometries, confining therefore the parameter space of realistic search 
for correlated circles as indicated, for example, by Mota~\emph{et  al.\/}%
~\cite{MGRT2003a}.

\section{Detectability of cosmic topology} 
\label{sec:detect}

In the previous sections we have assumed that the topology of the 
universe is detectable, and focussed our attention on strategies and 
methods to discover or even determine a possible nontrivial topology
of the universe.  In this section we shall examine the consequences of 
this underlying detectability assumption in the light of the current 
astro-cosmological observations which indicate that our universe is 
nearly flat ($\Omega_0 \approx 1$)~\cite{Tegmark-et-al2003}.
Although this near flatness of the universe does not preclude a
nontrivial topology it may push the smallest characteristic size of 
$M$ to a value larger than the observable horizon $\chi_{hor}$, making 
it difficult or even impossible to detect by using multiple images
of radiating sources (discrete cosmic objects or CMBR maps). 
The extent to which a nontrivial topology may or may not
be detected has been examined in locally flat~\cite{GR2003},
spherical~\cite{GRTav2001a,WeLeUz2003,MRT2003} or 
hyperbolic~\cite{GRTav2001b,GRTav2002,Weeks2002,MRT2003}
universes. The discussion below is based upon our articles%
~\cite{GRTav2001a,GRTav2001b,GRTav2002,MRT2003}, so we shall
focus on nearly (but not exactly) flat universes (for a study 
of detectability of flat topology see~~\cite{GR2003}).

The study of the detectability of a possible nontrivial
topology of the spatial sections $M$ requires topological 
typical scale which can be put into correspondence with 
observation survey depths.
A suitable characteristic size of $M$ is the so-called 
injectivity radius $r_{inj}(x)$ at $x \in M$, which is 
defined in terms of the length of the smallest closed 
geodesics that pass through $x$ as follows. 

A closed geodesic that passes through a point $x$ in a multiply
connected manifold $M$ is a segment of a geodesic in the
covering space $\widetilde{M}$ that joins two images of $x$.
Since any such pair of images are related by an isometry 
$g \in \Gamma$, the length of the closed geodesic associated 
to any fixed isometry $g$, and that passes through $x$, is 
given by the corresponding distance function 
\begin{equation}
\label{dist-function}
\delta_g(x) \equiv  d(x,gx) \; .
\end{equation}
The injectivity radius at $x$ then is defined by
\begin{equation}
\label{rinjx}
r_{inj}(x) = \frac{1}{2} \min_{g \in \widetilde{\Gamma}}\,
\{\,\delta_g(x)\,\} \; ,
\end{equation}
where $\widetilde{\Gamma}$ denotes the covering group without 
the identity map. Clearly, a sphere with radius $r < r_{inj}(x)$ 
and centered at $x$ lies inside a fundamental polyhedron of $M$.

For a specific survey depth $\chi_{obs}$ a topology is said to be 
undetectable by an observer at a point $x$ if $\chi_{obs} < r_{inj}(x)$, 
since in this case every image catalogued in the survey lies inside the 
fundamental polyhedron of $M$ centered at the observer's position $x$. In 
other words, there are no multiple images in the survey of depth 
$\chi_{obs}$, and therefore any method for the search of cosmic 
topology based on their existence will not work. 
If, otherwise, $\chi_{obs} > r_{inj}(x)$, then the topology is 
potentially detectable (or detectable in principle).

In a globally homogeneous manifold, the distance function for any 
covering isometry $g$ is constant. Therefore, the injectivity radius 
is constant throughout the whole space, and so if the topology is 
potentially detectable (or undetectable) by an observer at $x$, 
it is detectable (or undetectable) by any other observer at any 
other point in the same space. 
However, in globally inhomogeneous manifolds the injectivity radius 
varies from point to point, thus in general the detectability of 
cosmic topology depends on both the observer's position $x$ and 
survey depth. Nevertheless, for globally inhomogeneous manifolds 
one can define the global injectivity radius by 
\begin{equation}
\label{rinj}
r_{inj} =  \min_{x \in M} \{\,r_{inj}(x)\,\} \; ,
\end{equation}
and state an \emph{'absolute' undetectability} condition.
Indeed, for a specific survey depth $\chi_{obs}$
a topology is undetectable by any observer (located at any
point $x$) in the space provided that $r_{inj} > \chi_{obs}$.

Incidentally, we note that for globally inhomogeneous manifolds 
one can define the so-called injectivity profile $\mathcal{P}(r)$ 
of a manifold as the probability density that a point $x \in M$ 
has injectivity radius $r_{inj}(x) = r$. 
The quantity $\mathcal{P}(r)dr$ clearly provides the probability 
that $r_{inj}(x)$ lies between $r$ and $r+dr$, and so the 
injectivity profile curve is essentially a histogram depicting 
how much of a manifold's volume has a given injectivity radius (for 
more detail on this point see Weeks~\cite{Weeks2002}).
An important point is that the injectivity profile for non-flat 
manifolds of constant curvature is a topological invariant since 
these manifolds are rigid. 

In order to apply the above detectability of cosmic topology 
condition in the context of standard cosmology, we note that 
in non-flat RW metrics~(\ref{RWmetric}) , the scale factor 
$R(t)$ is identified with the curvature radius of the spatial 
section of the universe at time $t$, and thus $\chi$ can be 
interpreted as the distance of any point with coordinates 
$(\chi, \theta, \phi)$ to the origin (in the covering space) 
in units of curvature radius, which is a natural unit of length. 

To illustrate now the above condition for detectability
(undetectability) of cosmic topology, in the light of recent observations~\cite{WMAP,Tegmark-et-al2003} we assume 
that the matter content of the universe is well approximated 
by dust of density $\rho_m$ plus a cosmological constant 
$\Lambda$. In this cosmological setting the curvature radius 
$R_0$ of the spatial section is related to the total density 
parameter $\Omega_0$ through the equation 
\begin{equation}
\label{CurvRad}
R_0^2 = \frac{kc^2}{H_0^2(\Omega_0-1)} \; ,
\end{equation}
where $H_0$ is the Hubble constant, $k$ is the normalized 
spatial curvature of the RW metric~(\ref{RWmetric}), and
where here and in what follow the subscript $0$ denotes 
evaluation at present time $t_0$. Furthermore, in this
context the redshift-distance relation in units of the 
curvature radius, $R_0=R(t_0)$, reduces to 
\begin{equation}
\label{redshift-dist}
\chi(z) = \sqrt{|1-\Omega_0|} \int_1^{1+z} \hspace{-4mm}
\frac{dx}{\sqrt{x^3 \Omega_{m0} + x^2 (1- \Omega_0) + 
 \Omega_{\Lambda 0}}} \; ,
\end{equation}
where $\Omega_{m0}$ and $\Omega_{\Lambda 0}$ are, respectively,
the matter and the cosmological density parameters, and 
$\Omega_0 \equiv \Omega_{m0} + \Omega_{\Lambda 0}$.
For simplicity, on the left hand side of~(\ref{redshift-dist})
and in many places in the remainder of this article, we have 
left implicit the dependence of the function $\chi$ on the 
density components.

\begin{figure}[tbh]
\centerline{\def\epsfsize#1#2{0.5#1}\epsffile{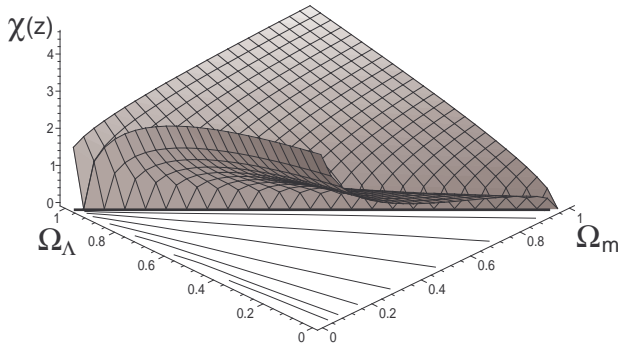}}
\caption{\label{fig:Bird} The behaviour of 
$\chi(\Omega_{m0},\Omega_{\Lambda0},z)$ for a fixed $z=1100$
as a function of the density parameters $\Omega_{\Lambda 0}$ and 
$\Omega_{m0}\,$.} 
\end{figure}

A first qualitative estimate of the constraints on detectability of 
cosmic topology from nearflatness can be obtained from the function
$\chi(\Omega_{m0},\Omega_{\Lambda0},z)\,$ given by~(\ref{redshift-dist})
for a fixed survey depth $z$. Figure~\ref{fig:Bird} clearly demonstrates 
the rapid way $\chi$ drops to zero in a narrow neighbourhood of the 
$\Omega_0 = 1$ line. This can be understood intuitively from~(\ref{CurvRad}), 
since the natural unit of length (the curvature radius $R_0$) goes 
to infinity as $\Omega_0 \to 1$, and therefore the depth $\chi$ (for 
any fixed $z$) of the observable universe becomes smaller in this 
limit.
{}From the observational point of view, this shows that the detection of 
the topology of the nearly flat universes becomes more and more difficult 
as $\Omega_0 \to 1$, a limiting value favoured by recent observations.
As a consequence, by using any method which relies on observations of 
repeated patterns the topology of an increasing number of nearly flat 
universes becomes undetectable in the light of the recent observations,
which indicate that $\Omega_0 \approx 1$.  

{}From the above discussion it is clear that cosmic topology 
may be undetectable for a given survey up to a depth $z_{max}$, 
but detectable if one uses a deeper survey. At present the deepest 
survey available corresponds to $z_{max}=z_{LSS}\approx 10^3$,
with associated depth $\chi(z_{LSS})$. So the most promising 
searches for cosmic topology through multiple images of radiating 
sources are based on CMBR.

To quantitatively illustrate the above features of the detectability
problem, we shall examine the detectability of cosmic topology of the 
first ten smallest (volume) hyperbolic universes.
 
To this end we shall take the following interval of the density 
parameters values consistent with current observations:
$\Omega_0 \in [0.99,1)$ and $\Omega_{\Lambda 0} \in [0.63,0.73]$.
In this hyperbolic sub-interval one can calculate the largest 
value of $\chi_{obs}(\Omega_{m0},\Omega_{\Lambda0},z)\,$ for the 
last scattering surface ($z=1100$), and compare with the injectivity 
radii $r_{inj}$ to decide upon detectability. From~(\ref{redshift-dist})
one obtains $\chi^{max}_{obs} = 0.337\,$.

\begin{table}[!ht]
\begin{center}
\begin{tabular}{|l|c|c|} \hline
$Manifold$ & $r_{inj}$ & {\sc cmbr}  \\ \hline \hline
\ m003(-3,1) & \ 0.292 \ & --- \\ 
\ m003(-2,3) & 0.289 & --- \\ 
\ m007(3,1)  & 0.416 & $U$ \\ 
\ m003(-4,3) & 0.287 & --- \\ 
\ m004(6,1)  & 0.240 & --- \\ 
\ m004(1,2)   & 0.183 & --- \\ 
\ m009(4,1)  & 0.397 & $U$  \\
\ m003(-3,4) & 0.182 & ---  \\
\ m003(-4,1) & 0.176 & ---  \\
\ m004(3,2)  & 0.181 & ---  \\
\hline
\end{tabular}
\caption{ \label{Tb:10smallest}
Restrictions on detectability of cosmic topology
for $\Omega_0\!\!=\!0.99 \;\mbox{with}\;
\Omega_{\Lambda 0} \in [0.63,0.73] $
for the first ten smallest known hyperbolic manifolds.
Here $U$ stands for undetectable topology with CBMR ($z_{max}=1100$), 
while the dash denotes detectable in principle.}
\end{center}
\end{table}

Table~\ref{Tb:10smallest} summarizes our results 
which have been refined upon and reconfirmed by Weeks~\cite{Weeks2002}. 
It makes explicit that there are undetectable topologies even if one 
uses CMBR. 

We note that similar results hold for spherical universes 
with values of the density parameters within the current observational 
bounds (for details see~\cite{GRTav2001a,WeLeUz2003,MRT2003}). This makes 
apparent that there exist nearly flat hyperbolic  and spherical universes 
with undetectable topologies for $\Omega_0 \approx 1$ favoured by 
recent observations. 

The most important outcome of the results discussed in this section 
is that, as indicated by recent observations (and suggested by inflationary 
scenarios) $\Omega_0$ is close (or very close) to one, then there are
both spherical and hyperbolic universe whose topologies are undetectable.
This motivates the development of new strategies and/or methods in the 
search for the topology of nearly flat universes, perhaps based on
the local physical effect of a possible nontrivial topology. In this regard see~\cite{ORT1994,BeGoReTe1998,RTT1998,GRTB2000,MuFaOp2001,MuFaOp2002}, for
example.

\section{Recent results and concluding remarks} 
\label{sec:news+remarks}

In this section we shall briefly discuss some recent results and
advances in the search for the shape of the universe, which have
not been treated in the previous sections. We also point out some 
problems, which we understand as important to be satisfactorily dealt 
with in order to make further progress in cosmic topology. 

One of the intriguing results from the analysis of WMAP data is
the considerably low value of the CMBR quadrupole and octopole moments, 
compared with that predicted by the infinite flat $\Lambda$CDM model.
Another noteworthy feature is that, according to WMAP 
data analysis by Tegmark \emph{et al.\/}, both the quadrupole and the 
octopole moments have a common preferred spatial axis along which 
the power is suppressed
\footnote{Incidentally, it was the fitting to the observed low values 
of the quadrupole and the octopole moments of the CMB temperature
fluctuations that motivated Poincar\'e dodecahedron space topology%
~\cite{Poincare}, which according to 'cirlces in the sky' plus WMAP 
analysis is excluded~\cite{CSSK2003}.
Nevertheless, the Poincar\'e dodecahedron space proposal was an
important step in cosmic topology to the extent that for the first 
time a possible nontrivial cosmic topology was tested against 
accurate CMBR data.}. 

This alignment of the low multipole moments has been suggested as
an indication of a direction along which a possible shortest closed 
geodesics (characteristic of multiply connected spaces) 
of the universe may be~\cite{OCTZH2003}. Motivated by this as well 
as the above anomalies, test using $S$-statistics~\cite{OCSS1996}
and matched circles furnished no evidence of a nontrivial
topology with diametrically opposed pairs of correlated circles%
~\cite{OCTZH2003}. It should be noticed, however, that these
results do no rule out most multiply connected universe models 
because $S$-statistics is a method sensitive only to Euclidean 
translations, while the search for circles in the sky, which
is, in principle, appropriate to detect any topology, was performed 
in a limited three-parameter version, which again is only suitable to 
detect translations. 

At a theoretical level, although strongly motivated by high precision 
data from WMAP, it has been shown that if a \emph{very} nearly flat
universe has a detectable nontrivial topology, then it will exhibit 
the generic local shape of (topologically) 
$\mathbb{R}^2 \times \mathbb{S}^1$%
\footnote{Or more rarely $\mathbb{R} \times \mathbb{T}^2\,$.},  
irrespective of its global shape~\cite{MGRT2003a}. In this case, 
{}from WMAP and SDSS the data analysis, which indicates that 
$\Omega_0 \approx 1$~\cite{Tegmark-et-al2003}, one has that if 
the universe has a detectable topology, it is very likely that it
has a preferred direction, which in turn is in agreement with 
the observed alignement of the quadrupole and octopole moments of 
the CMBR anisotropies. In this context, it is relevant to check
whether a similar alignment of higher order multipole ($\ell>3$)
takes place in order to reinforce a possible nontrivial local 
shape of our $3$--space. In this connection it is worth mentioning
that Hajian and Souradeep~\cite{HajianSouradeep2003a,HajianSouradeep2003b}  
have recently suggested a set of indicators $\kappa_{\ell}$ 
($\ell=1,2,3,...$) which for non-zero values indicate and quantify 
statistical anisotropy in a CMBR map.
Although $\kappa_{\ell}$ can be potentially used to discriminate
between different cosmic topology candidates, they give no
information about the directions along which the isotropy may
be violated, and therefore other indicators should be devised
to extract anisotropy directions from CBMR maps.

In ref.~\cite{MGRT2003a} it has also been shown that in a very 
nearly flat universe with detectable nontrivial topology, the 
observable (detectable) isometries will behave nearly like 
translations. Perhaps if one use Euclidean space to locally
approximate a nearly flat universe with detectable topology, the
detectable isometries can be approximated by Euclidean isometries,
and since these isometries are not translations, they have to be
screw motions. As a consequence, an approximate local shape of a
nearly flat universe with detectable topology would look like a 
\emph{twisted} cylinder, i.e. a flat manifold whose covering group 
is generated by a screw motion. Work toward a proof of this 
conjecture is being carried out by our research group.

Before closing this overview we mention that the study of the
topological signature (possibly) encoded in CMBR maps as
well as to what extent the cosmic topology CMBR detection methods
are robust against distinct observational effects such as, e.g., 
Suchs-Wolfe and the thickness of the LSS effects, will benefit 
greatly from accurate simulations of these maps in the context
of the FLRW models with multiply connected spatial sections. 
A first step in this direction has been achieved by Riazuelo 
\emph{et al.\/}~\cite{RULW2002}, with special emphasis on the 
effect of the topology in the suppression of the low
multipole moments. Along this line it is worth studying through 
computer-aided simulations the effect of a nontrivial cosmic 
topology on the nearly alignments of the quadrupole and the 
octopole moments (spatial axis along which the power is 
suppressed).

To conclude, cosmic topology is at present a very active research 
area with a number of important problems, ranging from how the 
characterization of the local shape of the universe may observationally 
be encoded in CMBR maps, to the development of more efficient 
computationally searches for matching circles, taking into 
account possible restrictions on the detectable isometries, and 
thereby confining the parameter space which realistic `circles in the 
sky' searches need to concentrate on. It is also of considerable 
interest the search for the statistical anisotropy one can 
expect from a universe with non-trivial space topology. 
Finally, it is important not to forget that there are almost 
flat (spherical and hyperbolic) universes, whose spatial topologies 
are undetectable in the light of current observations with the available
methods, and our universe can well have one of such topologies. In this 
case we have to devise new methods and strategies to detect the topology
of the universe.

\bigskip
\noindent{\bf Acknowledgments}
\medskip

We thank CNPq and FAPESP (contract 02/12328-6) for the grants under 
which this work was carried out. We also thank A.A.F. Teixeira 
and B. Mota for the reading of the manuscript and indication of 
relevant misprints and omissions.


\begin{thebibliography}{99}
\bibitem{WMAP} 
C.L. Bennett et al.\ , Astrophys.\ J.\ \textbf{583}, 1 (2003); \\ 
D.N. Spergel et al.\ , Astrophys.\ J.\ Suppl.\ \textbf{148}, 
175 (2003); \\
G. Hinshaw et al.\ , Astrophys.\ J.\ Suppl.\ \textbf{148}, 
135 (2003); \\
C.L. Bennett et al.\ , Astrophys.\ J.\ Suppl.\  \textbf{148}, 
1 (2003).

\bibitem{Ellis1971} G.F.R. Ellis, Gen.\ Rel.\ Grav.\ \textbf{2}, 7 
(1971).

\bibitem{SokolovShvartsman1974} D.D. Sokolov and V.F. Shvartsman, 
Sov.\ Phys.\ JETP \textbf{39}, 196 (1974). 

\bibitem{FangSato1985} L-Z. Fang and Sato, Gen.\ Rel.\ Grav.\ 
\textbf{17} 1117 (1985).

\bibitem{SokolovStarobinsky1975} D.D. Sokolov and A.A. Starobinsky, 
Sov.\ Astron.\  \textbf{19}, 629 (1975).

\bibitem{Gott1980} J.R. Gott, Mon.\ Not.\ R.\ Astron.\ Soc.\ 
\textbf{193}, 153 (1980).

\bibitem{Fagundes1983} H.V. Fagundes, Phys.\ Rev.\ Lett.\ \
\textbf{51}, 417 (1983).

\bibitem{FagundesWichoski1987} H.V. Fagundes and U.F. Wichoski, 
Astrophys.\ J.\, \textbf{322}, L52 (1987).

\bibitem{Gomero2003a} G.I. Gomero, Class.\ Quantum Grav.\ 
\textbf{20}, 4775 (2003).

\bibitem{LeLaLu1996} R. Lehoucq, M. Lachi\`{e}ze-Rey and J.-P. 
Luminet, Astron.\ Astrophys.\ \textbf{313}, 339 (1996).

\bibitem{LeLuUz1999} R. Lehoucq, J.-P. Luminet and  J.-P. Uzan,
Astron.\ Astrophys.\ \textbf{344}, 735 (1999).

\bibitem{FagGaus1998} H.V. Fagundes and E. Gausmann, \textsl{Cosmic
Crystallography in Compact Hyperbolic Universes}, 
astro-ph/9811368 (1998).

\bibitem{GTRB1998} G.I. Gomero, A.F.F. Teixeira, M.J. Rebou\c{c}as 
and  A. Bernui, Int.\ J.\ Mod. Phys.\ D \textbf{11}, 869 (2002). 
Also gr-qc/9811038 (1998). 

\bibitem{GRT2000a} G.I. Gomero, M.J. Rebou\c{c}as and  A.F.F. Teixeira,
Phys.\ Lett.\ A \textbf{275}, 355 (2000).

\bibitem{GRT2000b} G.I. Gomero, M.J. Rebou\c{c}as and A.F.F. Teixeira,
Int.\ J.\ Mod.\ Phys.\ D \textbf{9}, 687 (2000).

\bibitem{GRT2001a} G.I. Gomero, M.J. Rebou\c{c}as and  A.F.F. Teixeira,
Class.\ Quantum Grav.\ \textbf{18}, 1885 (2001).

\bibitem{BernuiTeixeira1999} A. Bernui and A.F.F. Teixeira,
\textsl{Cosmic crystallography: three multi-purpose functions\/},
gr-qc/9904180.

\bibitem{Reboucas2000} M.J. Rebou\c{c}as, Int.\ J.\ Mod.\ Phys.\ D 
\textbf{9}, 561 (2000).

\bibitem{UzLeLu1999} J.-P. Uzan, R. Lehoucq and J.-P. Luminet,
Astron.\ Astrophys.\ \textbf{351}, 766 (1999).

\bibitem{LeUzLu2000} R. Lehoucq, J.-P. Uzan and J.-P. Luminet, 
Astron.\ Astrophys.\ \textbf{363}, 1 (2000).

\bibitem{BeGoMoRe2004} A. Bernui, G.I. Gomero, B. Mota and M.J.
Rebou\c{c}as, in preparation (2004). A preliminar version was
presented by M.J. Rebou\c{c}as in the  section COT1: Topology 
of the Universe, of the \emph{X Marcel Grossmann Meeting\/} 
(July 20 -- 26, 2003).

\bibitem{Bernui2003} A. Bernui, private communication.

\bibitem{FagGaus1999} H.V. Fagundes and E. Gausmann,
Phys.\ Lett.\  A\ \textbf{261} 235 (1999).

\bibitem{RoukemaEdge1997} B.F. Roukema and A. Edge,
Mon.\ Not.\ R. Astron.\ Soc.\ \textbf{292}, 105 (1997).

\bibitem {Reviews} M. Lachi\`{e}ze-Rey and J.-P. Luminet, 
Phys.\ Rep.\ \textbf{254}, 135 (1995) ; \\
J.J. Levin, Phys.\ Rep.\ \textbf{365}, 251 (2002).

\bibitem{CSS1998} N.J. Cornish, D. Spergel and G. Starkman, Class.\ 
Quantum Grav.\ \textbf{15}, 2657 (1998).

\bibitem{CSSK2003} N.J. Cornish, D.N. Spergel, G.D. Starkman and E. 
Komatsu, \textsl{Constraining the Topology of the Universe\/}, 
astro-ph/0310233.

\bibitem{Poincare} J.-P. Luminet, J. Weeks, A. Riazuelo, R. Lehoucq 
and J.-P. Uzan, Nature \textbf{425}, 593 (2003).

\bibitem{Gomero2003b} G.I. Gomero, \textsl{`Circles in the Sky' 
in twisted cylinders\/}, astro-ph/0310749.

\bibitem{RiazUzLeLu2003} A. Riazuelo, J. Weeks, J.-P. Uzan and J.-P. Luminet,
\textsl{Cosmic microwave background anisotropies in multi-connected flat
spaces\/}, astro-ph/0311314.

\bibitem{MGRT2003a} B. Mota, G.I. Gomero, M.J. Reboucas and R. 
Tavakol, \textsl{What do very nearly flat detectable cosmic 
topologies look like?\/}, astro-ph/0309371.

\bibitem{Tegmark-et-al2003} M. Tegmark et al., \textsl{Cosmological 
parameters from SDSS and WMAP}, astro-ph/0310723.

\bibitem{GR2003} G.I. Gomero and M.J. Rebou\c{c}as, Phys.\ Lett.\ A  
\textbf{311}, 319 (2003).

\bibitem{GRTav2001a} G.I. Gomero, M.J. Rebou\c{c}as and  R. Tavakol,
Class.\ Quantum Grav.\ \textbf{18}, 4461 (2001).

\bibitem{WeLeUz2003}  J. Weeks, R. Lehoucq, J.-P. Uzan,
Class.\ Quant.\ Grav.\  \textbf{20}, 1529 (2003).

\bibitem{MRT2003} B. Mota, M.J. Rebou\c{c}as and  R. Tavakol,
Class.\ Quantum Grav.\ \textbf{20}, 4837 (2003).

\bibitem{GRTav2001b} G.I. Gomero, M.J. Rebou\c{c}as and  R. Tavakol,
Class.\ Quantum Grav.\ \textbf{18}, L145 (2001).

\bibitem{GRTav2002} G.I. Gomero, M.J. Rebou\c{c}as and  R. Tavakol,
Int.\ J.\ Mod.\ Phys.\ A \textbf{17}, 4261 (2002).

\bibitem{Weeks2002} J.R. Weeks, Mod.\ Phys.\ Lett.\  A \textbf{18}, 
2099 (2003).

\bibitem{ORT1994} W. Oliveira, M.J. Rebou\c{c}as and A.F.F. Teixeira,
Phys.\ Lett. A\ \textbf{188}, 125 (1994).

\bibitem{BeGoReTe1998} A. Bernui, G.I. Gomero, M.J. Rebou\c{c}as and 
A.F.F. Teixeira, Phys.\ Rev.\ D \textbf{57}, 4699 (1998).

\bibitem{RTT1998} M.J. Rebou\c{c}as, R. Tavakol and A.F.F. Teixeira,
Gen.\ Rel.\ Grav.\  \textbf{30}, 535 (1998).

\bibitem{GRTB2000} G.I. Gomero, M.J. Rebou\c{c}as, A.F.F. Teixeira
and A. Bernui, Int.\ J.\ Mod.\ Phys.\ A \textbf{15}, 4141 (2000).

\bibitem{MuFaOp2001} D. Muller, H.V. Fagundes and R. Opher,
Phys.\ Rev.\ D \textbf{63}, 123508 (2001).

\bibitem{MuFaOp2002} D. Muller, H.V. Fagundes and R. Opher
Phys.\ Rev.\  D \textbf{66}, 083507  (2002).

\bibitem{TOCH2003} M. Tegmark, A. de Oliveira--Costa and A.J.S. 
Hamilton, \textsl{A high resolution foreground cleaned CMB map from 
WMAP}, astro-ph/0302496.

\bibitem{FangHoujun1987} L.Z. Fang \& M. Houjun, Mod. Phys. Lett. 
A \textbf{2}, 229 (1987). 

\bibitem{Sokolov1993} I.Yu. Sokolov, JETP Lett. \textbf{57}, 617 
(1993). 

\bibitem{Starobinsky1993} A.A. Starobinsky, JETP Lett. \textbf{57}, 
622 (1993). 

\bibitem{OCTZH2003} A. de Oliveira--Costa, M. Tegmark, M. 
Zaldarriaga and A.J.S. Hamilton, \textsl{The significance of the 
largest scale CMB fluctuations in WMAP}, astro-ph/0307282.

\bibitem{OCSS1996} A. de Oliveira--Costa, G.F. Smoot \& A.A. 
Starobinsky, Astrophys.\ J.\, \textbf{468}, 457 (1996).

\bibitem{HajianSouradeep2003a} A. Hajian and T. Souradeep, 
\textsl{Statistical Isotropy of CMB and Cosmic Topology\/}, 
astro-ph/0301590.

\bibitem{HajianSouradeep2003b} A. Hajian and T. Souradeep,
\textsl{Measuring Statistical Isotropy of the CMB Anisotropy\/},
astro-ph/0308001.


\bibitem{RULW2002} A. Riazuelo, J.-P. Uzan, R. Lehoucq and J. 
Weeks, \textsl{Simulating Cosmic Microwave Background maps in 
multi-connected spaces}, astro-ph/0212223.

\end{thebibliography}
\end{document}